\documentclass[twocolumn,prl,aps,showpacs,superscriptaddress]{revtex4}

\usepackage[dvipdfm]{hyperref}

\usepackage{color}
\usepackage{soul}

\usepackage{graphicx}

\usepackage{amsmath}

\begin{document}

\title{Spin-Orbit Qubits of Rare-Earth-Metal Ions in Axially Symmetric Crystal Fields}



\author{S. Bertaina}
\affiliation{IM2NP, CNRS \& Universit\'e Paul C\'ezanne, Ave. Escadrille Normandie Niemen - Case 142 - 13397 Marseille Cedex 20, France}
\affiliation{Institut N\'eel, CNRS, BP 166, 38042 Grenoble Cedex 09,France}

\author{J.H. Shim}
\affiliation{Institut Nanosciences et Cryog\'enie, SCIB/LRM, CEA, 38054, Grenoble Cedex 09, France }

\author{S. Gambarelli}
\affiliation{Institut Nanosciences et Cryog\'enie, SCIB/LRM, CEA, 38054, Grenoble Cedex 09, France }

\author{B. Z. Malkin}
\affiliation{Kazan State University, Kazan 420008, Russian Federation}

\author{B. Barbara}
\affiliation{Institut N\'eel, CNRS, BP 166, 38042 Grenoble Cedex 09,France}
\affiliation{Institut Nanosciences et Cryog\'enie, SCIB/LRM, CEA, 38054, Grenoble Cedex 09, France }

\date{Submitted }

\begin{abstract}

Contrary to the well known spin qubits, rare-earth qubits are characterized by a strong influence of crystal field due to large spin-orbit coupling. At low temperature and in the presence of resonance microwaves, it is the magnetic moment of the crystal-field ground-state which nutates (for several $\mu$s) and the Rabi frequency $\Omega_R$ is anisotropic. Here, we present a study of the variations of $\Omega_R(\vec{H}_{0})$ with the magnitude and direction of the static magnetic field $\vec{H_{0}}$ for the odd $^{167}$Er isotope in a single crystal CaWO$_4$:Er$^{3+}$. The hyperfine interactions split the $\Omega_R(\vec{H}_{0})$ curve into eight different curves which are fitted numerically and described analytically. These "spin-orbit qubits" should allow detailed studies of decoherence mechanisms which become relevant at high temperature and open new ways for qubit addressing using properly oriented magnetic fields.

\end{abstract}


\maketitle

Worldwide studies on possible implementation of a spin-based quantum computer showed the significance of scalable single electron spin-qubits in e.g. quantum dots \cite{Oosterkamp1998, Petta2005, Koppens2006}. Are also interesting atomic and nuclear spins $1/2$ dispersed in solid state matrices with coherence times persisting at relatively high-temperature \cite{Dutt2007, Nellutla2007, Bertaina2009} or a new class of potentially scalable systems which recently emerged with the observation of Rabi oscillations \cite{Rabi1937} in the single molecular magnet V$_{15}$ \cite{Bertaina2008}, followed by electronic-spin coherence studies on Fe$_8$ \cite{Takahashi2009} and Fe$_4$ \cite{Schlegel2008}. Spin $1/2$ qubits are isotropic or quasi-isotropic and their Rabi frequency $\Omega_R=g\mu_Bh_{mw}/2h$ depends only on the amplitude of the linearly polarized microwave field
$h_{mw}$ ($h$ is the Planck constant, $\mu_B$ is the Bohr magneton and $g\sim 2$).

With rare-earths ions in crystals \cite{Bertaina2007}, the situation is very different.
Contrary to qubits with magnetic moments simply proportional to the spin (single electrons, $3d$ transition metal ions with zero or quenched orbital moment), the orbital magnetic moment of a rare earth ion is large and not reduced by the crystal-field, this last being much weaker than the spin-orbit coupling. In general case this leads to a strong magnetic anisotropy and, as it will be shown in this paper (see preliminary results in \cite{Bertaina2007}), to a strong dependence of the Rabi frequency on the directions of the microwave and static applied fields. The amplitude of the total angular moment $\vec{J}= \vec{L}+\vec{S}$ is generally much larger than 1/2 leading to magnetic moments as large as 10$\mu_{B}$ (Ho$^{3+}$)
allowing spin manipulations in low driving fields. For odd isotopes, hyperfine interactions of $4f$ electrons with the rare earth nuclear spin can be large enough so that nuclear spins are entangled with the total angular moment \cite{Giraud2001} and modify the Rabi frequency \cite{Bertaina2008}.

In this paper we concentrate on the anisotropy of the Rabi frequency of a single crystal CaWO$_4$:Er$^{3+}$, a system which presents an in-plane magnetic anisotropy favoring the observation of Rabi oscillations and several Erbium isotopes (odd and even) allowing the study of coherent dynamics with and without nuclear spin. In the first part we give a theoretical description (numerical and analytical) of what we may call "spin-orbit qubits" and in the second one we present the experimental results confirming quantitatively the theoretical predictions and allowing to understand the effect of strong spin-orbit coupling on Rabi oscillations. \\

The crystals of CaWO$_4$:0.05\% Er$^{3+}$ used in this study, obtained by the Chokralsky method, are characterized by a body centered tetragonal scheelite-type structure ($I4_1/a$ space group) with the lattice constants $a=b=0.524$ nm and $c=1.137$ nm \cite{Hazen1985}. The Er$^{3+}$ ions substitute for the Ca$^{2+}$ ions inside the eight-fold oxygen surroundings, charge excess being compensated by supplementary substitutions of Na$^+$ ions in the crystal. The Hamiltonian operating in the space of states of the ground $^{4}I_{15/2}$ multiplet of a single Er$^{3+ }$ ion in CaWO$_4$ ($S_4$ point symmetry) is given by:

\begin{equation}\label{eq:1}
    H= H_{cf}+ H_{hf}+ H_{Z}
\end{equation}

where $H_{cf}$ is the Hamiltonian of a single rare-earth ion, reduced by the host-matrix crystal-field and $4f$ symmetry:

\begin{eqnarray}\label{eq:2}
H_{cf}&=&\alpha_JB_2^0O_2^0+\beta_J(B_4^0O_4^0+B_4^4O_4^4+B_4^{-4}O_4^{-4})\nonumber\\			 &&+\gamma_J(B_6^0O_6^0+B_6^{4}O_6^{4}+B_6^{-4}O_6^{-4})
\end{eqnarray}

The $O_l^m$ are the Stevens equivalent operators, $\alpha_J$, $\beta_J$, $\gamma_J$ the Stevens coefficients \cite{Stevens1952} and the $B_l^m$ are the crystal-field parameters.
For CaWO$_4$:Er$^{3+}$, $B_2^0$=231, $B_4^0$=-90,$B_4^4$=$\pm$852, $B_4^{-4}$=0, $B_6^0$=-0.6, $B_6^4$=$\pm$396, $B_6^{-4}$=$\pm$75 cm$^{-1}$
in the Cartesian system of coordinates with the $z$-axis parallel to the $c$-axis. The second term in (1), $H_{hf}=A_J\vec{J}.\vec{I}$, is the magnetic hyperfine interaction for the $^{167}$Er isotope ($I=7/2$, 22.9\% abundance, hyperfine constant $A_{J}$ = -125MHz \cite{Bernal1971}), quadrupolar couplings are negligible. The third term in (1), $H_{Z}=g_J\mu_B\vec{J}.\vec{H_0}$ ($g_J=6/5$ is the Land\'e factor of Er$^{3+}$), is the static Zeeman interaction. We neglect here dipole-dipole interactions between the highly diluted Er$^{3+}$ ions and super-hyperfine interactions with nuclear magnetic moments of the $^{183}$W isotope. These interactions induce negligible changes of Rabi frequencies, but may be principal sources of decoherence \cite{Prokof'ev2000,Zvezdin1998,Hanson2008,Dobrovitski2009}. The crystal-field of CaWO$_4$ leads to strong in-plane anisotropy of Er$^{3+}$ ($a-b$ plane)\cite{Bernal1971} and effective g-factors  $g_{||}=g^c$=1.2,  $g_\perp =g^a$= $g^b$= 8.4 \cite{Kurkin1970}.
The Rabi frequency can be calculated either in the laboratory frame (LF) by computing the amplitude of probability of transition induced by the microwave field $h_{mw}$ perpendicular to the static field $H_{0}$ or in the rotating frame (RF) by computing the splitting at avoided level crossing.

In the LF, we start with the Hamiltonian \eqref{eq:1} of a single $^{167}$Er$^{3+}$ ion. By diagonalization of the 128 dimension Hilbert space electro-nuclear Hamiltonian (J=15/2$\otimes$I=7/2), we calculate the Rabi frequencies as a function of the direction of the static field $\vec{H_0}$ using the expression $\Omega_R =\mid g_J\mu_B\vec{J}_{pk}.\vec{h}_{mw}\mid/2h$ where $\vec{J}_{pk}= \langle\phi_p(\vec{H}_0)|\vec{J}|\phi_k(\vec{H}_0)\rangle$ is a vector varying with the static field $\vec{H}_{0}$ and $\phi_{p,k}(\vec{H}_{0})$ are the wave functions resulting from the diagonalization of \eqref{eq:1}. In general, the anisotropy of $\Omega_R $ depends on both the directions of $\vec{H}_{0}$ and $\vec{h}_{mw}$. These expressions for spin-orbit qubits in a crystal-field generalize the one of simple spin qubits, $\Omega_R=g^\alpha\mu_B h^\alpha_{mw}/2h$ \cite{Schweiger2001}  ($g^\alpha$ is the effective g-factor in the direction $\alpha$ of the electromagnetic wave polarization $h^\alpha_{mw}$).

\begin{figure}

\centering

\includegraphics[bb=64 15 801 560,width=\columnwidth]{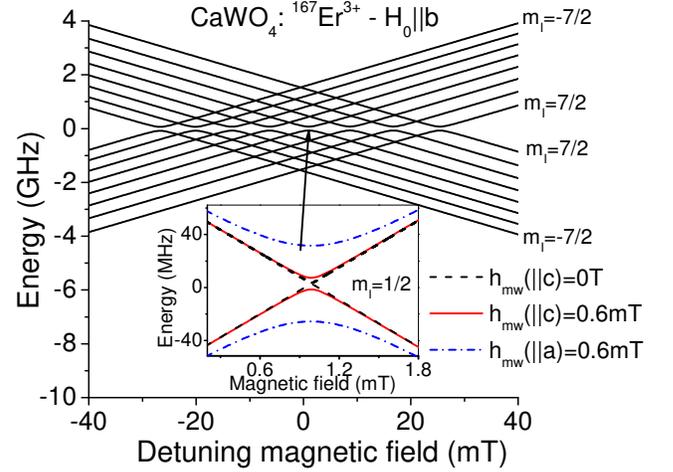}

\caption{(color online) Energy spectrum of $^{167}$Er$^{3+}$ with $I$=7/2 calculated numerically in the rotating frame. The detuning field is, by definition, the difference between the applied field $H_0$ and the resonant field of the $I$=0 isotope. The Rabi frequencies are equal to the splittings at avoided level crossings (dressed states). Allowed EPR transitions ($\Delta m_I=0$) are gapped and their anisotropy for different directions of the microwave polarization is shown in the inset.}

\label{fig:1}

\end{figure}

In the RF, we use the time-dependent effective Hamiltonian of the ground states doublet :

\begin{equation}\label{eq:3}
    H(t)=\mu_B\vec{H}_0[g]\vec{S}+\vec{I}[A]\vec{S}+\mu_B\vec{h}_{mw}[g]\vec{S}\cos(2\pi f t)
\end{equation}
$f$ is the microwave frequency, $[g]$ the g-tensor, $[A]= [g]A_{J}/g_{J}$ the hyperfine tensor. The NMR paradigm applied to EPR \cite{Leuenberger2001,Leuenberger2003} allows to transform \eqref{eq:3} into a time independent effective Hamiltonian. The Rabi frequencies are then given by avoided levels splitting. The Rabi frequencies of the $I$=7/2 $^{167}$Er$^{3+}$ isotope are derived from the RF energy spectrum obtained numerically for $H_0||a=b$, Fig.\ref{fig:1}. The effective electro-nuclear spin-states $|1/2, m_I \rangle$ and  $|-1/2, m_I\rangle$ coupled by the electromagnetic wave ("dressed states") form several avoided level crossings of which splitting is $h\Omega_R$ (decreasing rapidly when the microwave field rotates from the easy $a-b$ plane to the hard $c$-axis). In the case of the $I=0$ isotope, when $\vec{H}_0$ makes an angle $\phi$ with $c$-axis in the $x$-$z$ plane and $h_{mw}$ is in the same plane, the result can be analytical and given by:

\begin{equation}\label{eq:4}
    \Omega_R(\phi)=\mu_B h_{mw}\frac{g_\perp g_{||}}{2h  g(\phi)}
\end{equation}

where  $g(\phi)=\sqrt{g_{||}^2\cos^2\phi+g_{\perp}^2\sin^2\phi}$. Much more extended analytical results will be given in a forthcoming publication \cite{tobepublished} showing that both LF and RF method give, in all cases, quite the same results. In the following we will use the LF method which is more convenient for precise calculations.

 The EPR measurements have been performed using a conventional Bruker EPR spectrometer E580. This spectrometer works at both, continuous wave (CW) or time resolved (TR) mode in the X band. All experiments were carried out at the frequency of $f=$ 9.7 GHz. The detection of the signal was performed using a cavity working with the TE$_{011}$ mode with $Q$-factor of about 4000 in CW mode (perfect coupled cavity) and 200 in TR mode (over-coupled cavity). The sample-cavity ensemble is surrounded by a $^4$He flux cryostat and the temperature was controlled by an Oxford instrument ITC form 2.5 to 300 K. Two pulse sequences have been used i) the usual "spin-echo sequence" which strongly reduces the decoherence associated with inhomogeneous CW line-width, and ii) the Rabi oscillations technique consisting in a spin-echo sequence preceded by an excitation pulse. This technique allows to obtain the time evolution of the averaged magnetization proportional to $\langle S_z(t)\rangle$, i.e. the Rabi oscillations if the timescale is such that coherence wins over relaxation (here the $z-$axis $||\vec{H_{0}}$).

\begin{figure}

\centering

\includegraphics[bb=35 57 768 617,width=\columnwidth]{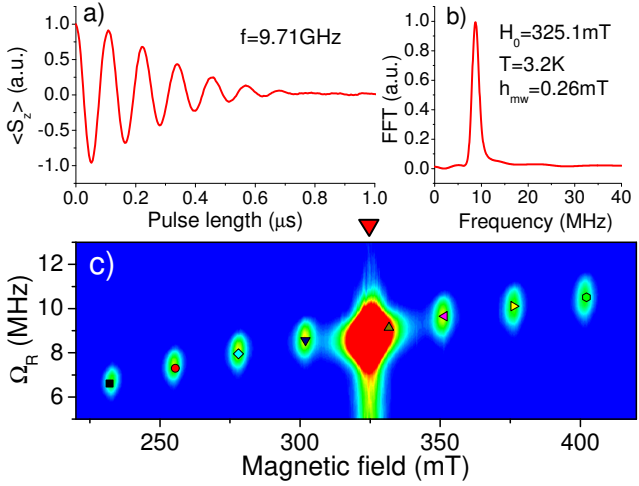}

\caption{(color online) (a) Rabi oscillation recorded on the nuclear-spin free Er$^{3+}$ isotope in CaWO$_4$ ($H_{0}=325.1mT$, $h_{mw}=0.13mT$) and (b) its Fourier transform. (c) Contour plot of the Rabi frequency distribution measured in the $I=7/2$ isotope while $H_0$ tilted by 12$^\circ$ from the c-axis is swept from 220mT to 440mT (colored iso-magnetization scale increases from blue to red). The eight Rabi frequencies with smaller intensity correspond to the transitions $(1/2,m_{I})\rightarrow (-1/2,m_{I})$ of the $I$=7/2 isotope ($\Delta m_I$ = 0) and the single one with larger intensity to the $I$=0 isotope. Calculated $\Omega_R$ are represented by symbols going from black square ($m_I=-7/2$ ) to green circle ( $m_I=7/2$). }

\label{fig:2}

\end{figure}

Fig.\ref{fig:2}(a) and (b) give an example  of measured Rabi oscillation and its Fourier transform whereas Fig.\ref{fig:2}(c) shows the eight electro-nuclear transitions of the $I=7/2$ isotope, plus the single one of the nuclear-spin free isotope. In this experiments $\vec{H}_0$ was tilted by $\phi=12^\circ$ from the $c$-axis in order to resolve the eight electro-nuclear transitions 
. The symbols correspond to the  $\Omega_R$= 6.61, 7.30, 7.95, 8.57, 9.14, 9.65, 10.1, 10.5MHz calculated from diagonalization of the electro-nuclear Hamiltonian \eqref{eq:1} (LF method) using the published crystal-field parameters and hyperfine constant only \cite{Bernal1971}. The RF expression \eqref{eq:4} generalized to the case $I\neq$0 and linear in $A_{J}$ give the values $\Omega_R$= 6.63, 7.22, 7.82, 8.42, 9.02, 9.6, 10.21, 10.80MHz \cite{tobepublished}. The small differences between the two methods are caused by the neglect of higher terms in $A_{J}$. Both sets of values agree very well with experimental data.

\begin{figure}
\centering
\includegraphics[bb=46 17 742 543,width=\columnwidth]{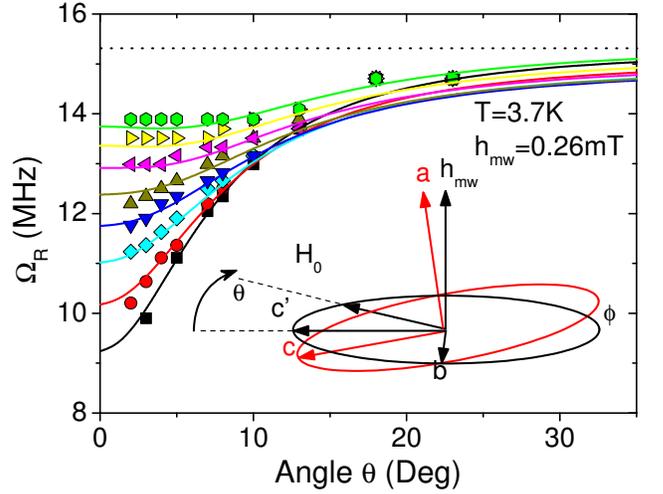}

\caption{(color online) Evolution of $\Omega_R$ when the static field $H_0$ rotates by an angle $\theta$ around the microwave field $h_{mw}$. Colored continuous lines are computed numerically. Inset: geometry of the experiment where the sample defined by its axes $a,b,c$ is tilted by a small angle $\phi\sim 7^\circ$ around the $b$=$b'$-axis. The dotted line represents the conventional equation $\Omega_R=g^\alpha\mu_B h^\alpha_{mw}/2h$ \cite{Schweiger2001}.}
\label{fig:3}
\end{figure}

\begin{figure}

\centering

\includegraphics[bb=46 17 742 543,width=\columnwidth]{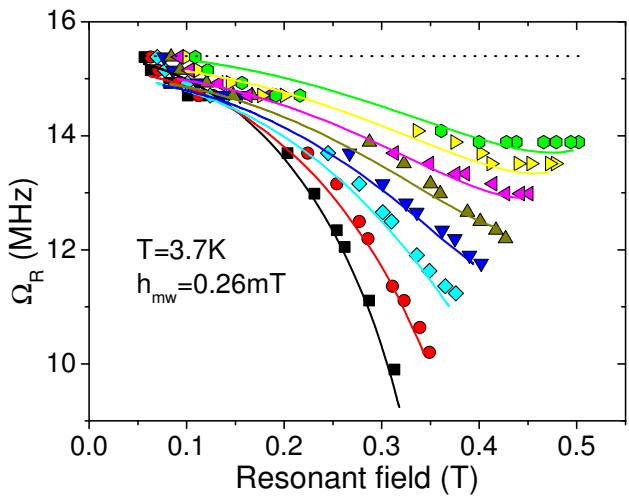}

\caption{(color online) Evolution of $\Omega_R$ with $\Omega_R$ vs $\parallel\vec{H}_0\parallel$ while the resonant field is rotated. The symbols represent the same data and the lines are calculated numerically and analytically as in Fig. 3. The dotted line represents the conventional equation $\Omega_R=g^\alpha\mu_B h^\alpha_{mw}/2h$ \cite{Schweiger2001}.}

\label{fig:4}

\end{figure}

Fig. \ref{fig:3} shows the evolution of the Rabi frequency of each electro-nuclear transition when the crystal rotates from $\theta=$0$^\circ$ ($\vec{H}_0||c'$) to $\theta=$90$^\circ$ ($\vec{H}_0||b$) (see inset).
Above $\theta\approx 30^\circ$ (not shown) all the Rabi frequencies $\Omega_{R}(m_{I},\theta)$ tend asymptotically to the value $\Omega_R\sim$ 15MHz predicted by the usual equation $\Omega_R=g^\alpha\mu_B h^\alpha_{mw}/2h$ \cite{Schweiger2001} independently of $m_{I}$.
At $\theta<30^\circ$ these quasi-degenerated frequencies $\Omega_{R}(m_{I},\theta)$ split off in a way strongly dependent on the algebraic value of $m_{I}$. It is the small tilt angle $\phi$ which allows to resolve the transitions with different $m_{I}$ (i.e. for $\phi=$0,  $\Omega_R$ does not depend on $m_{I}$).
Non-magnetic isotopes show a similar variation laying between the blue and green triangles with $\Lambda_R=\Omega_R(H_0||c') /\Omega_R(H_0\perp c')\approx$ 0.77. In these experiments a rotation of the resonance field $\vec{H}_{0}$ implies a change of its amplitude in order to keep at resonance, and this rises the question to know whether the observed variations of $\Omega_{R}(m_{I},\theta)$  are mainly due to changes in direction or amplitude of the vector field $\vec{H}_0$. In order to answer this question, we plot in Fig. \ref{fig:4} the evolution of $\Omega_R$ vs $\parallel\vec{H}_0\parallel$. The large dispersion observed on $\Omega_R$ at given $\parallel\vec{H}_0\parallel$ definitively confirms that it is, as expected, a directional effect. All the measured Rabi frequencies of Fig. \ref{fig:3} and Fig. \ref{fig:4} are very well reproduced by our crystal-field model without any fitting parameter using only published values of parameters, as described above. In particular, the ratio $\Lambda_R$ varies from 0.58 to 0.90, whereas, in absence of orbital contribution, $\Lambda_R(m_I)=1$ for any value of $m_I$.

Interestingly, if the coordinate system is such that the microwave field $\vec{h}_{mw}$ has only one non-zero component ($h^\alpha_{mw}$), the scalar-product equation
 $\Omega_R = \mid g_J\mu_B\vec{J}.\vec{h}_{mw}\mid/2h$ takes the simpler form :
$\Omega_R(\vec{H}_0,\vec{h}_{mw}) = G(\alpha,\beta)\mu_Bh^\alpha_{mw}/2h$ where $G(\alpha,\beta)
=g_J |\langle\phi_1(H_0^\beta)|J_\alpha|\phi_2(H_0^\beta)\rangle |$
depends on both directions $\alpha$ and $\beta$ of the static and microwave fields ($\vec{\alpha}.\vec{\beta} = 0$). This expression traduces the fact that, when $\vec{H}_0$ rotates around $\vec{h}_{mw}$ the crystal-field/Zeeman competition varies leading renormalization of wave functions entailing variations of $G(\alpha,\beta)$ , i.e. of $\Omega_R$. As mentioned above, this is true only if the crystal is tilted so that $\vec{h}_{mw}$ is not along a crystallographic axis ($\phi\neq$0, inset Fig.\ref{fig:3}). In the opposite case where $\vec{h}_{mw}$ is along a crystallographic axis ($a$-axis in Fig.\ref{fig:3}) $\Omega_R$ is isotropic and follows the usual expression $\Omega_R=g^\alpha\mu_B h^\alpha_{mw}/2h$ \cite{Schweiger2001} represented by a dashed line in Fig.\ref{fig:3} and \ref{fig:4}. This renormalization of wave functions appears clearly in the analytical expressions for the non-magnetic and magnetic isotopes: $\Omega_R$ depends on both $g_{\parallel}$ and $g_{\perp}$ if $\phi\neq$0 whereas for $\phi=$0 it depends on $g_{\perp}$ only, and $\Omega_{R}(m_{I})$ associated with the different $m_{I}$ are clearly different from each others unless $\phi=$0 or $\pi$/2 \cite{tobepublished} .

In conclusion, we have shown with the example of Er$^{3+}$ ions diluted in the single-crystalline host matrix CaWO$_4$, that crystal-field deeply modifies the coherent quantum dynamics of what we call "spin-orbit qubits". In particular, it introduces dramatically large variations of the Rabi frequency $\Omega_{R}$ when the static or/and dynamical field ($\vec{H}_{0}$ or $\vec{h}_{mw}$) deviate from crystallographic axes. This effect, directly connected with modifications of the wave functions of resonant states is amplified or depressed by the hyperfine interaction of the odd isotope ($I=$7/2) which removes the eight-fold degeneracy of $\Omega_{R}$ by increasing or decreasing it when $m_{I}>$0 or $m_{I}<$0. All the experimental data, showing a rich pattern of coherent oscillations, are interpreted quantitatively without requiring any fitting parameter giving a clear picture of this new type of crystal-field dependent qubits. These qubits are interesting for several reasons (i), the coherent nutation of the ground-state magnetic moment deriving from crystal-field effects acting on $\vec{J}=\vec{L}+\vec{S}$ (and $\vec{J}+\vec{I}$ with odd isotopes) is associated with , not yet well studied, symmetry and temperature-dependent spin-lattice decoherence mechanism; in particular, the transfer of quantum dynamics from the spin-bath to the nutating system is probably partial implying the existence of residual decoherence \cite{Prokof'ev2000,Zvezdin1998,Hanson2008,Dobrovitski2009}.
(ii) Despite significant spin-phonon coupling, relatively long living coherence is observed ($\approx 50\mu s$ at 2.5K in CaWO$_4:$Er$^{3+}$) showing that the coupling with crystal-field environment is not redhibitory. (iii) The magnetic moment generally much larger than 1/2 allows spin manipulations in low driving field-vectors (amplitude and direction). (iv) RE qubits inserted in a semi-conducting film, should be scalable with selective addressing (application of weak local field pulses created by nano-line current adding algebraically to uniform static field) and couplings (controlled carrier injection through a gate voltage).

\begin{acknowledgements}

We thank for financial support the former INTAS-99-01839 contract in which the crystals of Er:CaWO$_4$ were synthesized, the CEA, and the European Network of Exellence MAGMANET.

\end{acknowledgements}

\bibliographystyle{apsrev}

\begin{thebibliography}{24}
\bibitem[{tob()}]{tobepublished}
\bibinfo{journal}{Details of calculations will be published elsewhere}.
\end{thebibliography}

\end{document}